\documentclass[11pt,a4paper]{article}
 
\usepackage[margin=1in]{geometry}
\usepackage{amsmath}
\usepackage{amssymb}
\usepackage{graphicx}
\usepackage{bm}
\usepackage{mathtools}
\usepackage{authblk}
\usepackage{cite}
\usepackage{hyperref}
\usepackage{multibib}
\usepackage{xr}
\newcites{supp}{References}

\title{\Large Resolving topological obstructions to vectorial structured field control}
\date{}
\author[1*]{An Aloysius Wang}
\author[1]{Yuxi Cai}
\author[1]{Yifei Ma}
\author[1]{Patrick S Salter}
\author[1*]{Chao He}
\affil[1]{Department of Engineering Science, University of Oxford, Parks Road, Oxford, OX1 3PJ, UK}
\affil[*]{Corresponding authors: aloysius.wang@gmail.com, chao.he@eng.ox.ac.uk}

\begin{document}
\maketitle
{\bf The use of structured matter, such as optical retarders, for vectorial control is a well-established and widely employed technique in modern optics, and has driven continued advances in the manipulation of complex, spatially varying vectorial fields. However, achieving arbitrary field conversion typically requires the use of cascaded elements, as intrinsic physical and fabrication constraints fundamentally limit individual devices to a restricted subset of transformations. This results in an overall continuous transformation potentially failing to be continuous at the level of the parameters of the cascade, leading to detrimental engineering consequences such as the introduction of complex, discontinuous aberrations that disrupt important topological properties of the underlying matter field. In this work, we establish a novel mathematical framework for analyzing the topological difficulties that emerge in the decomposition of an overall transformation into individual layers, and for determining the minimal depth required to overcome them. The strategy introduced provides a general pathway for optimizing designs for vectorial field control and matter field generation, with particular significance for the manipulation of topological phases in optical polarization fields, such as Stokes skyrmions, where continuity is of vital importance.}\\

The application of structured fields has enabled breakthroughs in diverse areas \cite{He2022}, including microscopy \cite{he_polarisation_2021, Deng2023_H_E_birefringence}, adaptive optics \cite{He2023VectorialAdaptiveOptics, He2020_VectorialAdaptiveOptics, Zhang2023_OptLett_6136, Shen2022_PolarizationAberrations}, sensing \cite{Ma2025OpticalSkyrmions, Cheng2025MetrologyWithATwist}, optical communication \cite{Willner2021OrbitalAngularMomentumForCommunications}, photonic computing \cite{Wang2025PerturbationResilient}, and quantum optics \cite{ornelas_non-local_2024, ornelas2024topologicalrejectionnoisenonlocal}. In many of these applications, the manipulation of phase and polarization follows a two-step design process that first involves determining a desired transformation using Jones \cite{lu_homogeneous_1994} or Mueller calculus \cite{Jorge2022}, and second, realizing it through a polarization-sensitive device, typically an optical retarder \cite{BornWolf1999PrinciplesOfOptics}. The reason for the latter is that spatially varying optical retarders, such as liquid-crystal devices, metasurfaces, and waveplate arrays, can be fabricated through a variety of different approaches and at variety of different scales and degrees of reconfigurability, making them highly versatile platforms for vectorial control \cite{Ji2023MetasurfaceDesignQuantumOptics, yu_light_2011, balthasar_mueller_metasurface_2017, yu_broadband_2012}. 

However, the core physical principle behind many such optical retarders is the use of anisotropic elements that impart different phase delays to orthogonal linear polarizations, with control achieved either by tuning the orientation of the anisotropy, as in Pancharatnam-Berry (PB) phase metasurfaces \cite{Chen2012DualPolarityPlasmonicMetalens}, or by varying the phase difference, as in liquid-crystal devices \cite{Yang2023_LC_SLM_Review}. In both cases, the possible Jones or Mueller matrices achievable by a single layer lie on an $S^1$ submanifold of $\mathrm{SU}(2)$ and $\mathrm{SO}(3)$, respectively, and is generally insufficient to realize arbitrary control. As such, the complex manipulation of structured fields often relies on cascades of lower-functionality devices \cite{Converter} arranged such that their combined action forms an elliptical retarder \cite{Zhang2025Elliptical} and spans a higher-dimensional subset of all possible Jones or Mueller matrices \cite{he2023universal}, thereby providing the expressibility required to perform desired transformations.

In such cascaded structures, an important design consideration is the decomposition of an overall transformation into the parameters of the cascade. However, due to topological differences between the parameter space of the target field and that of the cascade, ensuring that an overall continuous transformation remains continuous at the level of the cascade parameters is not always straightforward. More importantly, when this condition is not met, two significant practical issues arise---one engineering in nature and the other topological.

\begin{figure}[!t]
    \centering
    \includegraphics[width=\textwidth]{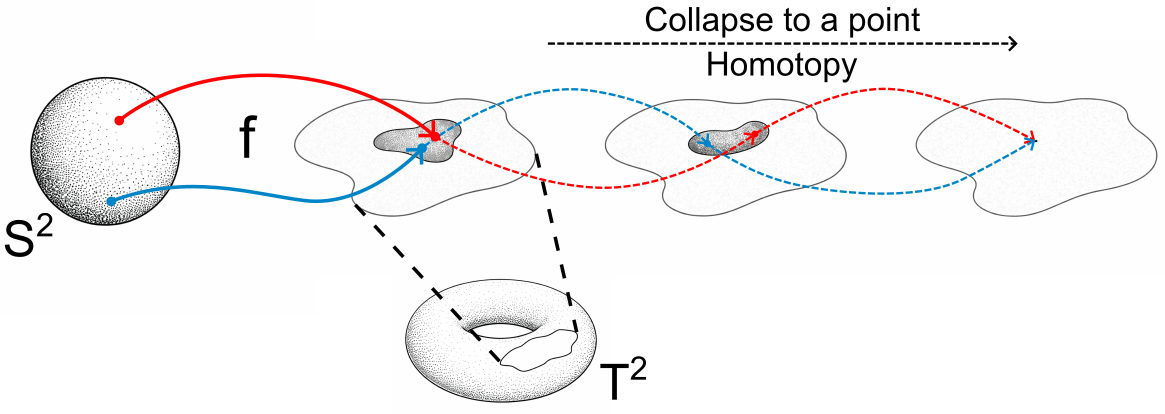}
    \caption{{\bf Concept.} A pictorial representation illustrating the vanishing of $\pi_2(T^2)$, where $T^2 = S^1 \times S^1$ denotes the standard torus. Consider the image of a map $f \colon S^2 \to T^2$, depicted as a blob above. For any such $f$, there exists a homotopy that continuously collapses the image to a point, implying that any two maps from $S^2$ into $T^2$ are homotopic. The figure also illustrates, in red and blue, the trajectories of two representative points throughout this process. }
\label{fig:concept}
\end{figure}

Before expanding on these issues in more detail, we first explain how they arise and highlight two important use cases in which they are commonly encountered, namely light and matter field generation. Light field generation typically involves converting a uniform input field into a specified spatially varying target Stokes or Jones vector field, and is important for vector beam generation \cite{wang2024topological, Han2013VectorialOpticalFieldGenerator, Rong2014GenerationArbitraryVectorBeams}, structured illumination \cite{chen2023superresolution}, and related applications \cite{he_polarisation_2021}. Taking polarization as our main example, suppose the target field is given by $\mathcal{S}\colon \mathbb{R}^2 \longrightarrow S^2$. Viewing a cascade of length $n$ as a continuous map 
$p\colon (S^1)^n \longrightarrow S^2$, a na\"ive approach to designing the parameters of each layer of the cascade is to pick out a solution for each polarization state $f\colon S^2 \longrightarrow (S^1)^n$, namely
\begin{equation*}
    p \circ f = \mathrm{id}_{S^2},
\end{equation*}
so that $f \circ \mathcal{S}$ yields the required parameter distribution. Since $p$ is typically smooth, a simple dimensional argument via Sard's theorem shows that $n \geq 2$ for such a map $f$ to exist. Indeed, this has been explicitly demonstrated using a cascade of two spatial light modulators (SLMs) aligned at 45\textsuperscript{$\circ$} to each other \cite{wang2024topological}. 

As alluded to above, a more subtle question is whether $f$ can be taken to be continuous, in which case the parameter distribution $f \circ \mathcal{S}$ inherits the continuity properties of $\mathcal{S}$. This is relevant from a practical perspective for several reasons, most notably that if $f$ is not continuous at a given polarization state, the corresponding cascade parameters can vary wildly when generating such states, leading to a system that is highly prone to errors (see Methods 1). Moreover, discontinuities in the cascade parameters can lead to artefacts such as phase and intensity aberrations that distort the quality of the generated field, and may result in increased fabrication difficulty and unwanted edge effects. A more detailed explanation of these challenges, including examples of real-world systems in which they arise, is provided in Methods 1.

Beyond these engineering considerations, topological effects also come into play. For example, from the perspective of optical Stokes skyrmions \cite{wang2024topological, shen_optical_2023, Tsesses2018, Gao2020}, which have recently attracted interest for applications in optical communications \cite{Shen2023} and photonic computing \cite{Wang2025PerturbationResilient}, continuity constraints play a central role, not only in ensuring that topological indices are well defined \cite{shen_optical_2023}, but also in enabling robust manipulation of the skyrmion number \cite{Wang2025PerturbationResilient}.

Crucially, for many applications ranging from beam analysis, such as full Poincar\'e beam polarimetry \cite{Beckley2010FullPoincareBeams, he2023universal}, to beam generation, these engineering and topological issues become unavoidable, since the target field often contains every possible polarization state. However, there is a general topological obstruction to the existence of $f$, namely the second homotopy group: $\pi_2(S^2) \cong \mathbb{Z}$, whereas $\pi_2((S^1)^n) \cong 0$, as illustrated in Fig.\ \ref{fig:concept}. Therefore, if we consider the induced map in homotopy, we have that 
\begin{equation*}
    p_\ast \circ f_\ast = \mathrm{id}_{\pi_2(S^2)}
\end{equation*}
which clearly cannot be true. This implies that, under the na\"ive approach, no matter the cascade length, there exists no arrangement of spatially uniform intermediate components for which one is guaranteed continuous parameters for all continuous target fields. A similar but more involved argument applies to the cases of Jones vector field generation and matter field generation (see Discussion).

In order to overcome the topological obstruction arising from a fundamental mismatch in homotopy groups, it becomes necessary to continuously select a distinct solution at each point in space. Using SLM cascades as an illustrative example, we demonstrate in this work how such issues can be rigorously analyzed, and we show that, for the application of skyrmion beam generation, a continuous decomposition can be achieved with three SLMs provided that a certain \v{C}ech cohomology class constructed via $\mathcal{S}$ is trivial, whereas a four-SLM cascade is, in general, sufficient to eliminate any such obstructions. We go on to demonstrate the importance of continuous parameter distributions for enabling a system to exhibit topological properties, such as the ability to modulate the skyrmion numbers of incident fields. Lastly, we present results extending the theory above to matter field generation and show that a similar obstruction in \v{C}ech cohomology arises in a four-SLM cascade, and discuss extensions to cascades of length five and greater, as well as to full vectorial control of phase, polarization, and intensity. The strategy for beam and matter field generation presented represents a shift in perspective on field engineering, treating it as a global problem constrained by topology rather than a pointwise one, and opens the door to advanced topological field engineering and manipulation. 

\section*{Main}

We attack the problem of continuous phase decomposition by considering SLM cascades of increasing length, starting from a minimal two-SLM configuration for which arbitrary polarization field generation is possible \cite{wang2024topological}, and exploring the topological obstructions present at each stage. In the case of two SLMs, one possible implementation is based on the following decomposition 
\begin{equation*}
    \mathcal{S} = \begin{pmatrix}
        s_1\\s_2\\s_3
    \end{pmatrix} = \begin{pmatrix}
        \sin\hat{\chi} \sin \hat{\phi} \\ \cos \hat{\chi} \\ \sin\hat{\chi}\cos\hat{\phi}
    \end{pmatrix} = \begin{pmatrix}
        \cos \hat{\phi} & 0 & \sin \hat{\phi} \\ 0 & 1 & 0 \\ -\sin \hat{\phi} & 0 & \cos\hat{\phi}
    \end{pmatrix}\begin{pmatrix}
        1 & 0 & 0 \\ 0 & \cos\hat{\chi} & -\sin\hat{\chi} \\ 0 & \sin \hat{\chi} & \cos \hat{\chi}
    \end{pmatrix}\begin{pmatrix}
        0 \\ 1 \\ 0
    \end{pmatrix},
\end{equation*}
which physically corresponds to $45^\circ$-polarized light passing through a two-SLM cascade, where the first SLM introduces a retardance of $\hat{\chi}$ and has its fast axis aligned at $0^\circ$, and the second introduces a retardance of $\hat{\phi}$ with its fast axis aligned at $45^\circ$. 

Here, we treat the cascade parameter, namely retardance, as an $\mathbb{R}$-valued quantity, and it is also the variable directly controlled by voltage in practice. However, the corresponding Mueller matrix depends only on the $S^1$-valued quantity $\chi = e^{\mathrm{i}\hat{\chi}}$, whose non-trivial topology leads to the previously mentioned difficulties in achieving a continuous decomposition. Note also that, because $\mathbb{R}^2$ is contractible, standard results in covering space theory ensure that any continuous $\chi \colon \mathbb{R}^2 \to S^1$ lifts to a continuous function $\hat{\chi} \colon \mathbb{R}^2 \to \mathbb{R}$. Therefore, topological obstructions in the $\mathbb{R}$-valued setting and the $S^1$-valued setting are mathematically equivalent, and using retardance as the 
variable of choice is simply a matter of convenience. There are, however, differences from a practical perspective, namely implementing a continuous $\chi$ through a continuous voltage distribution requires the SLM to have a sufficiently large dynamic range to accommodate the unwrapped field $\hat{\chi}$. Any phase discontinuities arising from an insufficient dynamic range are therefore implementation artefacts rather than true topological obstructions, and their removal becomes a matter of engineering.

With the decomposition above, we obtain
\begin{equation*}
    \hat{\chi} = \arccos(s_2), \qquad \hat{\phi} = \begin{cases}
        \operatorname{atan2}(s_1, s_3), & s_2 \neq \pm 1, \\
\text{arbitrary}, & s_2 = \pm 1,
    \end{cases}
\end{equation*}
which shows that $\hat{\phi}$ may fail to be continuous at points where $\mathcal{S} = \pm(0, 1, 0)$, corresponding to $\pm 45^\circ$ linearly polarized light. 

\begin{figure}[!t]
    \centering
    \includegraphics[width=\textwidth]{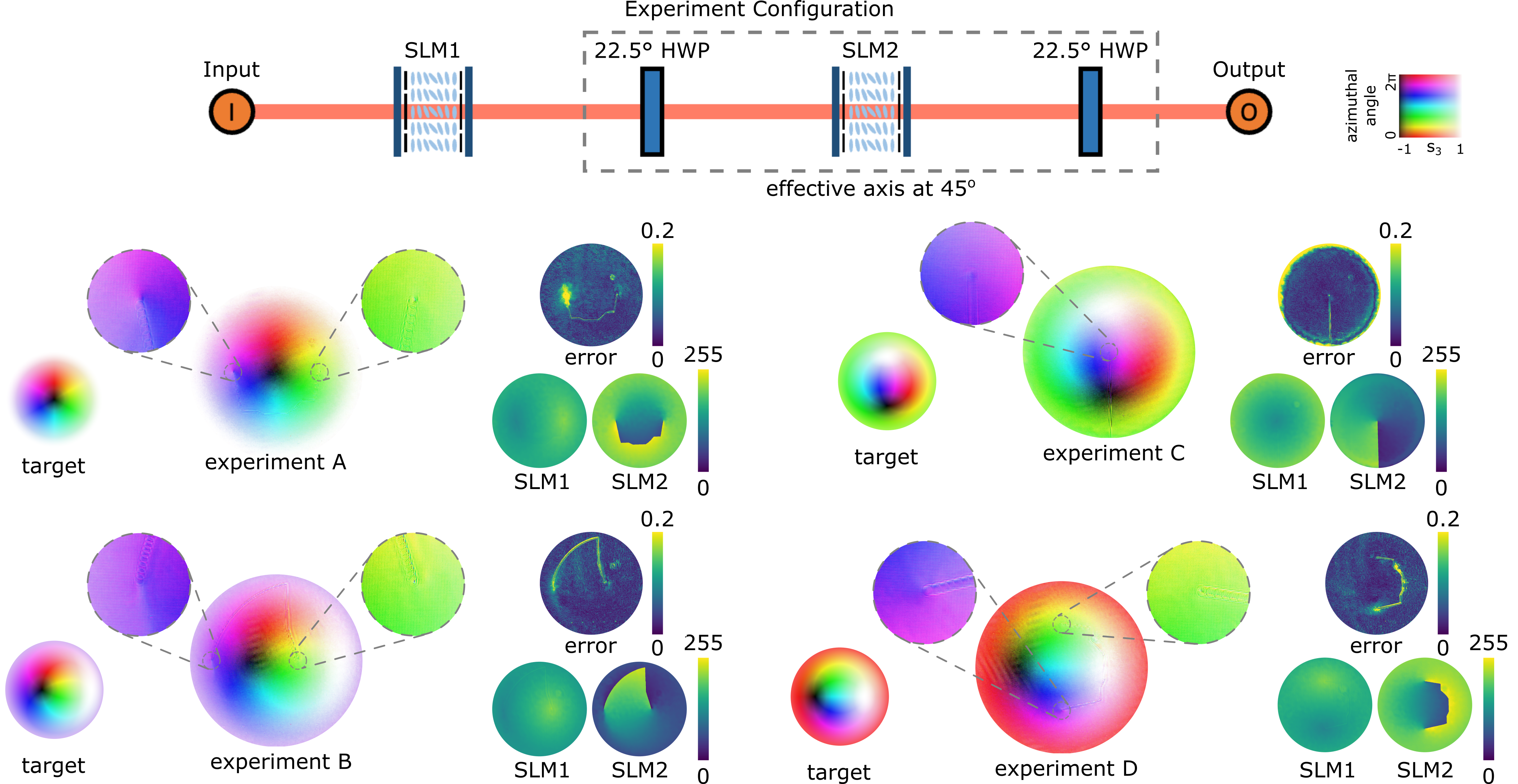}
    \caption{{\bf Experimental results (2 SLM cascade).} Target and experimentally measured Stokes fields of four different degree-1 skyrmions, with the corresponding experimental configuration shown above. Throughout this paper, Stokes fields are depicted using color to represent the azimuthal angle on the Poincar\'{e} sphere and saturation to represent height \cite{shen_optical_2023}. The experimental set-up is drawn with {\bf I} representing the incident field, which is linearly polarized at $45^\circ$, and {\bf O} representing the output field. Note that the second SLM is sandwiched between half-wave plates (HWPs) whose fast axes are aligned at $22.5^\circ$, such that the composite system acts as a linear retarder with its fast axis aligned at $45^\circ$. Insets highlighting the output Stokes field near $\pm 45^\circ$ linearly polarized light are shown, in which a line artefact is clearly visible. The corresponding phase patterns (given in levels) applied to each SLM are also shown. A clear phase singularity is observed on the second SLM at points where the target field is $\pm 45^\circ$ linearly polarized, in agreement with the established theory. Lastly, the $\ell^2$-error distribution between the generated and target fields is presented, where the errors arising from phase discontinuities are clearly visible.}
\label{fig:2SLM}
\end{figure}

In the context of skyrmion generation, the surjectivity of nonzero-degree skyrmions necessarily leads to such discontinuities, providing one of the key motivations for studying cascades of greater length. For example, Fig.\ \ref{fig:2SLM} presents experimentally measured Stokes fields \cite{Azzam16, Azzam78, Yuxi} of four degree-1 skyrmions generated using a cascade of two SLMs arranged in the configuration described above, where the discontinuity issue is clearly evident. In each case, a singularity appears in the phase of the second SLM whenever the target Stokes field is $\pm 45^\circ$ linearly polarized, corresponding to the green and purple regions in our plots. Consequently, a clearly observable line artefact arises in the resulting polarization state, as highlighted by the insets in the figure. The figure also includes the $\ell^2$-error distribution between the experimentally measured field and the target field, which demonstrates that the phase discontinuity significantly amplifies error in beam generation. Of particular note is experiment C, where the field approaches $45^\circ$ linear polarization along its boundary. As a result, the entire region near the boundary exhibits increased error due to the discontinuity of the inverting map at that polarization state.

Note that an \textit{in situ} feedback correction strategy is employed during beam generation to account for errors in converting between SLM voltage level and phase, as well as for possible misalignments and aberrations within the system. However, the feedback correction cannot remove the phase singularity, further 
reinforcing that it is a fundamental limitation of the system. As evidenced by the error distribution in experiment C, the generated polarization state becomes increasingly insensitive to the phase applied on the second SLM as it approaches $\pm 45^\circ$ linear polarization. Consequently, there are effectively fewer degrees of freedom available for manipulation near this state, and the error near the boundary of experiment C cannot be  suppressed even with feedback control.

Summarizing the results above, for a two-SLM cascade the inverting map from polarization state to SLM phase is fixed for all polarization states except $\pm 45^\circ$ linear polarization. This lack of flexibility prevents the resolution of topological obstructions in beam generation, making a discontinuous phase decomposition unavoidable.
To overcome this, additional SLMs must be introduced into the cascade, as further degrees of freedom are required to expand the solution space for field generation, thereby allowing access to different solution branches that offer a potential route to resolving these topological issues. In what follows, we begin with a general analysis of cascades of arbitrary length before specializing to cascades of length three and four.

More precisely, consider an arbitrary-length cascade obtained by extending the two-SLM structure above through the inclusion of additional SLMs with retardances $\hat{\psi}_1, \ldots, \hat{\psi}_n$ oriented at various angles, together with various waveplates, so that the overall Mueller matrix of the transformation is given by some $M(\hat{\psi}_1, \ldots, \hat{\psi}_n) \in \mathrm{SO}(3)$. Solving for $\hat{\chi}$ and $\hat{\phi}$ in terms of $\hat{\psi}_1,\ldots, \hat{\psi}_n$, we may define vectors $u, v, w$ implicitly via
\begin{equation*}
    M^{T}\mathcal{S} = \begin{pmatrix}
        u^T\mathcal{S} \\ v ^T\mathcal{S} \\ w^T\mathcal{S}
    \end{pmatrix} = \begin{pmatrix}
        \sin\hat{\chi} \sin \hat{\phi} \\ \cos \hat{\chi} \\ \sin\hat{\chi}\cos\hat{\phi}
    \end{pmatrix}
\end{equation*}
from which it is clear that a continuous branch of $\hat{\phi}$ may be chosen provided that we can select $\hat{\psi}_1, \ldots, \hat{\psi}_n$ satisfying the following condition.
\begin{equation}\label{eq: main}
\text{\textbf{Criterion for arbitrary-length cascades: }}
\mathcal{S}\neq \pm u\times w.
\end{equation}
Determining whether this criterion holds underpins our analysis of topological obstructions for all subsequent cascades of length greater than two.

Specializing to a three SLM cascade, in the scenario where a single SLM with fast axis aligned at $0^\circ$ is added, the right-hand side of Eq.\ \ref{eq: main} evaluates to
\begin{equation*}
    u \times w = \begin{pmatrix}
        0 \\ -\cos \hat{\psi}_1 \\ -\sin \hat{\psi}_1
    \end{pmatrix}.
\end{equation*}
Thus, the general criterion above easily reduces to finding a continuous $\hat{\psi}_1$ such that $s_2 + \mathrm{i}s_3 \neq \pm \psi_1 = \pm e^{\mathrm{i}\hat{\psi}_1}$ at every point in space. Note, however, that such a construction is not always possible and depends on the topology of the set
\begin{equation} \label{eq: sigma}
    \Sigma \coloneqq \{(x,y) \colon s_1(x,y)=0\}.
\end{equation}
An explicit example illustrating this is presented in Methods 2.

More generally, the existence of $\psi_1$ depends entirely on whether the map $\tau \colon \Sigma \longrightarrow S^1$, as defined in Methods 2, can be lifted to $\mathbb{R}$ (Supplementary Note 1.1). However, determining whether such a lift exists is in general difficult, since $\Sigma$ may be arbitrarily pathological. Indeed, one can construct a corresponding continuous field $\mathcal{S}$ for any closed set $\Sigma$, including sets that are not locally path-connected and for which standard covering space theory does not apply. As such, a proper characterization of the problem relies on \v{C}ech cohomology, namely
\begin{center}
\noindent\textbf{Criterion for 3-SLM cascade:} $\tau$ lifts $\Leftrightarrow [\tau]=0\in\check{H}^1(\Sigma;\mathbb{Z}),$
\end{center}
where $[\tau]$ is defined in Supplementary Note 1.1.

Fig.\ \ref{fig:3SLM} presents experimental results demonstrating beam generation with a three-SLM cascade, from which the effect of the topology of the set $\Sigma$ can be visually observed. Note that in each case presented, $\Sigma$ is sufficiently regular such that the \v{C}ech cohomology groups are naturally isomorphic to the corresponding singular cohomology groups, the latter of which can be computed by standard means. More concretely, let $r$ be a normalized radial coordinate such that $r = 1$ corresponds to the boundary of the field. In experiments A--C, for $r < 1$, the set $\Sigma \cap B_r(0)$ is just a straight line, and $\check{H}^1(\Sigma \cap B_r(0);\mathbb{Z}) \cong H^1(\Sigma \cap B_r(0);\mathbb{Z}) = 0$. Therefore, for $r < 1$, it is possible to find a continuous phase distribution on the third SLM that guarantees a continuous phase decomposition throughout $B_r(0)$, as demonstrated by our experiments. However, at $r=1$, one has $\check{H}^1(\Sigma;\mathbb{Z}) \cong H^1(\Sigma;\mathbb{Z}) \cong \mathbb{Z}^2$, and it is clear that any lift of $\tau$ across the straight line connecting the outer circular boundary will differ by $2\pi$ at its two endpoints, thereby obstructing a lift to all of $\Sigma$. As a result, for experiments A--C, the addition of a third SLM allows the phase discontinuity originally present in the two-SLM cascade to be pushed to the boundary, and the phases on each SLM can be made continuous for $r < 1$.

\begin{figure}[!t]
    \centering
    \includegraphics[width=\textwidth]{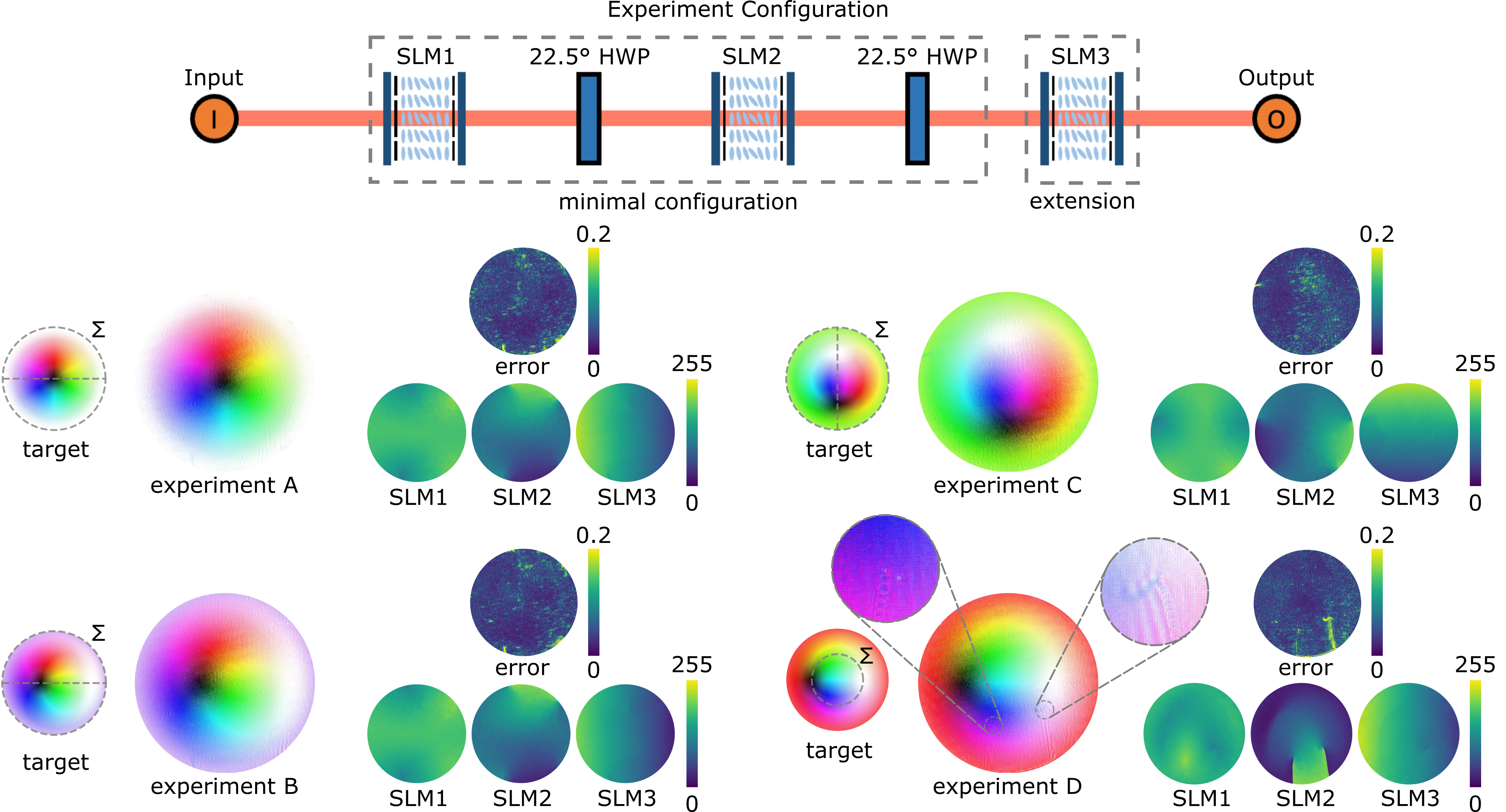}
    \caption{{\bf Experimental results (3 SLM cascade).} Target and experimentally measured Stokes fields of four different degree-1 skyrmions, with the corresponding experimental configuration shown above. The distinguished set $\Sigma$, as defined in Eq.\ \ref{eq: sigma}, is highlighted by dashed gray lines on the target field. The corresponding phase patterns on each SLM, along with the $\ell^2$-error distributions between the generated and target fields, are also shown. For experiment D, insets highlighting the Stokes field near phase discontinuities are included, where a line artefact is clearly visible.}
\label{fig:3SLM}
\end{figure}

In experiment D, $\Sigma$ is given by the circle $r = 0.5$, and represents the special case $\Sigma \cong S^1$, $\deg \tau \neq 0$ discussed in Methods 2. As such, despite the addition of the third SLM, there is no way of removing the phase singularity on the second SLM, which appears, as predicted, along the circle $r = 0.5$. Notice also from the figure that the effects of the phase discontinuity in experiment D are directly reflected in the amplified error of the generated field along the discontinuity, as in the case with two SLMs. This once again highlights the benefits of having a continuous phase distribution.

Summarizing the results above, in the case of a three-SLM cascade the addition of an extra SLM provides additional degrees of freedom that can resolve discontinuous phase decomposition in certain situations. Moreover, whether continuous phase decomposition can be achieved is entirely characterized by an easily computable cohomology class, which provides a practical criterion for assessing target fields. 

\begin{figure}[!t]
    \centering
    \includegraphics[width=\textwidth]{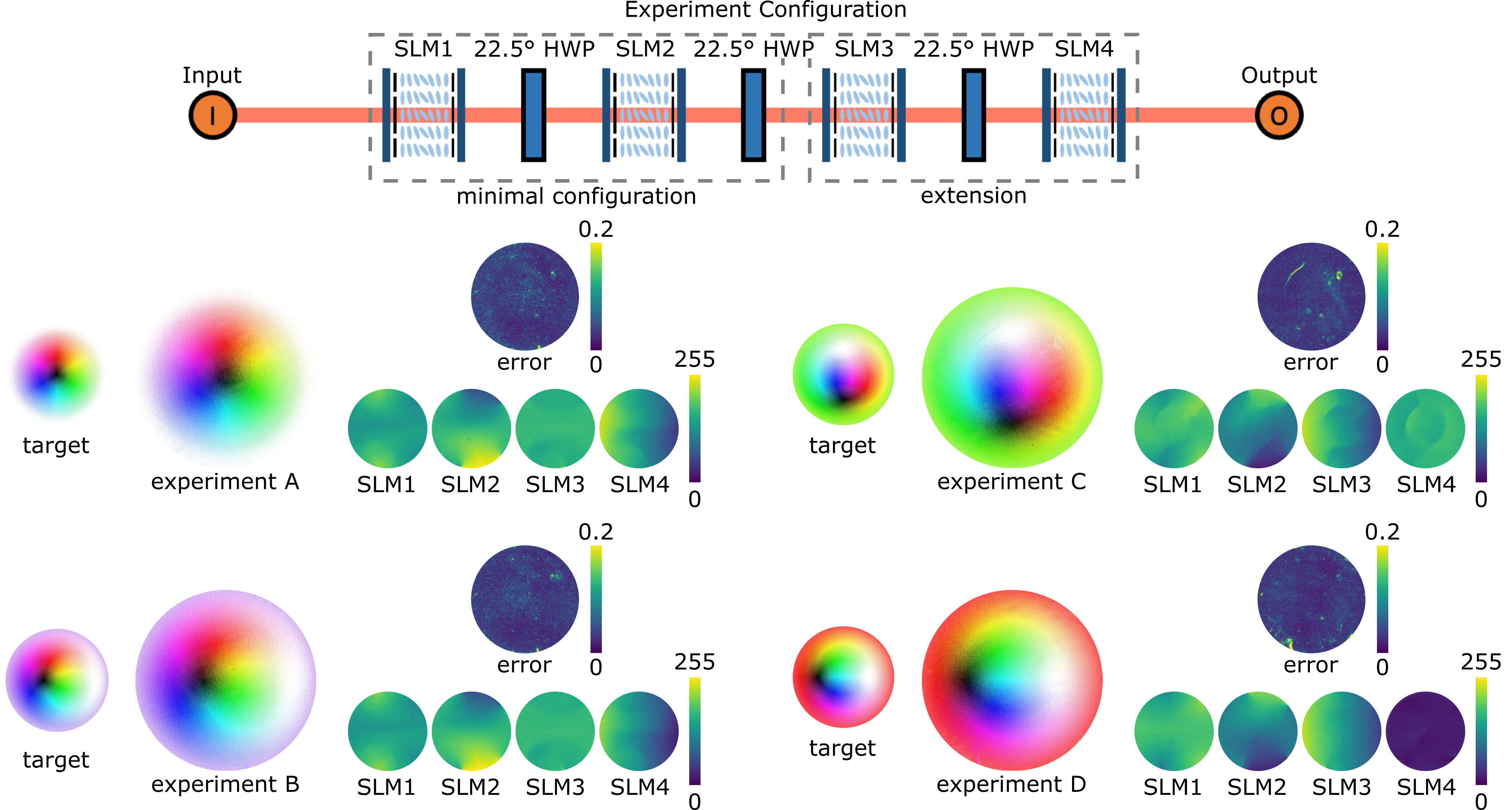}
    \caption{{\bf Experimental results (4 SLM cascade).} Target and experimentally measured Stokes fields of four different degree-1 skyrmions, with the corresponding experimental configuration shown above. The corresponding phase patterns on each SLM, along with the $\ell^2$-error distributions between the generated and target fields, are also shown.}
\label{fig:4SLM}
\end{figure}

Lastly, we consider the case of four SLMs. Here, we analyze the situation in which a half-wave plate, with its fast axis aligned at $22.5^\circ$, is added, followed by a final SLM whose fast axis is aligned at $0^\circ$. In this case, the right-hand side of criterion \ref{eq: main} evaluates to
\begin{equation*}
    \tilde{\mathcal{S}} \coloneqq u\times w = \begin{pmatrix}
        -\cos \hat{\psi}_1 \\ - \sin \hat{\psi}_1\sin \hat{\psi}_2 \\ \sin\hat{\psi}_1 \cos \hat{\psi}_2
    \end{pmatrix}.
\end{equation*}

Noting that there exists an obvious continuous inverting map $S^2 \backslash \{(1,0,0), (-1,0,0)\} \longrightarrow S^1 \times S^1$ for $e^{\mathrm{i}\hat{\psi}_1}$ and $e^{\mathrm{i}\hat{\psi}_2}$, a sufficient criterion for continuous decomposition is thus to find a field $\tilde{S}$ that differs everywhere from $\pm \mathcal{S}$ and avoids the points $\pm(1,0,0)$. Thus, we have
\begin{center}
\noindent\textbf{Criterion for 4-SLM cascade:} $\exists \tilde{\mathcal{S}}$ such that $\tilde{\mathcal{S}}(x,y) \neq \left\{ \pm \mathcal{S}(x,y),\, \pm(1,0,0) \right\}$ for all $ (x,y)\in\mathbb{R}^2$.
\end{center}
As expected, this is a significantly weaker condition than in the three-SLM case. A detailed discussion of the obstructions to the existence of such a field $\tilde{\mathcal{S}}$ is given in Supplementary Note 1.2.

Fig.\ \ref{fig:4SLM} presents experimental results for the cascade proposed above, from which it is clear that the addition of the fourth SLM resolves the phase discontinuity present in experiment D when fewer SLMs are used. Note that for our choice of skyrmions, which are rotations of standard N\'eel type skyrmions and have a relatively simple polarization structure, uniformly fixing either $\hat{\psi}_1$ or $\hat{\psi}_2$ carefully can reduce the situation to one that is mathematically equivalent to the favorable three-SLM case. In this case, one of the SLMs can in fact be replaced by a waveplate, thereby reducing the overall system complexity. Nonetheless, for more complex fields, the inclusion of a fourth SLM provides a weaker lifting condition that enables the resolution of topological obstructions beyond those accessible with a homogeneous wave plate.

Apart from overcoming engineering difficulties and minimizing error, a further key aspect of continuous decomposition is the topological character of the resulting matter fields, whereby the action of the medium on any incident field reduces to an operation on skyrmion numbers \cite{Wang2025PerturbationResilient}. More specifically, a continuous decomposition of parameters implies that the transformation of any incident field through the medium can be regarded as a homotopy; thus, the change in skyrmion number through such a cascade depends only on the boundary conditions of the incident field. This, in turn, implies that the design process presented above, while originally formulated for polarization field generation, actually provides a general strategy for constructing matter fields that achieve topologically protected arithmetic with skyrmion numbers, extending key results presented in \cite{Wang2025PerturbationResilient}. 

\begin{figure}[!t]
    \centering
    \includegraphics[width=\textwidth]{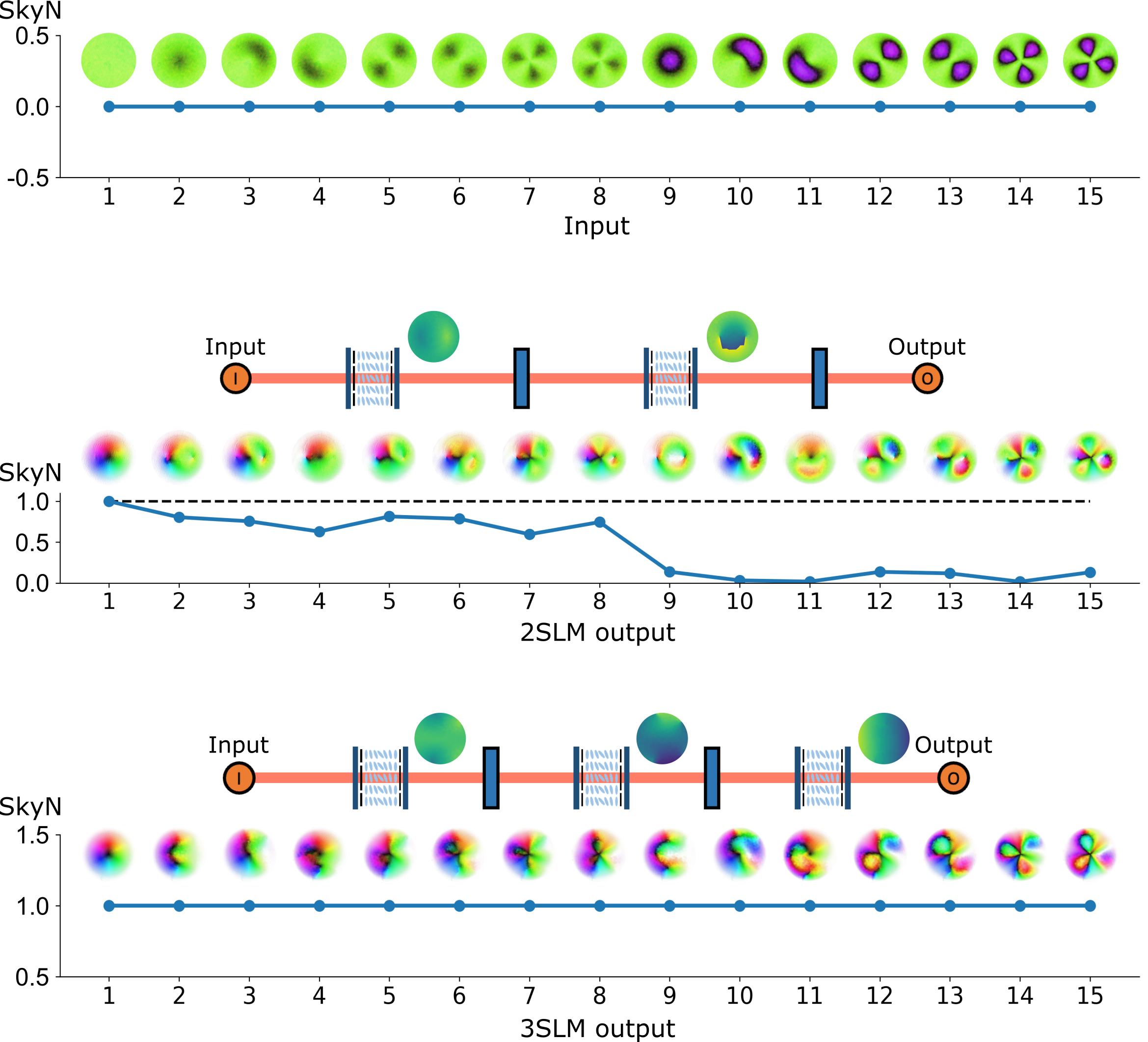}
    \caption{{\bf Experimental results (skyrmion).} Experimentally measured Stokes fields for different input beams passing through a beam generator designed to produce a standard N\'eel-type skyrmion from uniformly $45^\circ$ linearly polarized light, using cascades of two and three SLMs. Relevant experimental set-ups and SLM phase distributions are provided for completeness. The corresponding numerically computed skyrmion numbers are also presented, demonstrating instability of the skyrmion number when continuous parameter decomposition is not achieved, and stable skyrmion numbers when it is.}
\label{fig:skyrmion}
\end{figure}

We further emphasize that the skyrmion generation strategies developed in this work differ from traditional holography-based techniques relying on the superposition of Laguerre-Gaussian modes \cite{Zeng2025UltraHighOrderSkyrmions,Shen2022OpticalSkyrmions, Gao2020} in that they exhibit topological robustness. This is because, in our generation scheme, local perturbations to the incident and underlying matter fields induce local perturbations in the output field, thereby ensuring topological protection against disturbances whose support does not intersect the field boundary \cite{wang2024topological}. By contrast, in holographic methods, perturbations to the incident and underlying matter fields induce perturbations that are localized in $k$-space rather than in real space. As a result, the uncertainty principle implies that disturbances, regardless of their physical location, affect the boundary of the output field, leading to a loss of topological protection \cite{Liu2022}.

To demonstrate this topological aspect of continuous decomposition, Fig.\ \ref{fig:skyrmion} presents experimentally measured Stokes fields together with numerically computed skyrmion numbers for different incident fields with identical boundary conditions, propagating through different cascades. Two cascades are considered: one consisting of two SLMs and the other of three. The phase profiles for each cascade are designed to produce a standard N\'eel-type skyrmion when the incident light is uniformly $45^\circ$ polarized, corresponding to experiment A in Figs.\ \ref{fig:2SLM} and \ref{fig:3SLM}, respectively. In the two-SLM case, the decomposition is not continuous for the reasons discussed in the main text, and the output skyrmion number varies unpredictably with the input, as evidenced by the figure. By contrast, in the three-SLM case, a continuous parameter decomposition is achieved, resulting in a skyrmion number that remains stable and fixed at 1 for all inputs. Skyrmion numbers are computed using the Gaussian process regression technique described in \cite{wang2024generalizedskyrmions}. Taken together, these results support the theoretical prediction that continuous parameter decomposition yields stable changes in skyrmion numbers, demonstrating the practical relevance of our underlying framework for manipulating topological phases of polarization states.

\section*{Discussion}

Lastly, we turn to matter field generation directly. This is the second key use case introduced earlier and typically involves designing a cascade whose overall Jones or Mueller matrix realizes a prescribed spatially varying field. Such problems arise in applications including vectorial adaptive optics \cite{Hu2021ArbitraryComplexRetarders,He2023VectorialAdaptiveOptics,Ma2024Vectorial}, high-dimensional information storage \cite{chopsticks}, and arbitrary-to-arbitrary field conversion \cite{hu_arbitrary_2020}. In this case, the cascade can likewise be viewed as a continuous map into $\mathrm{SU}(2)$ or $\mathrm{SO}(3)$, respectively. However, the na\"\i ve approach fails once again, this time due to a different topological obstruction, namely a discrepancy in the third homotopy groups: $\pi_3(\mathrm{SU}(2)) \cong \mathbb{Z}$ and $\pi_3(\mathrm{SO}(3)) \cong \mathbb{Z}$, whereas $\pi_3\big((S^1)^n\big) \cong 0$. As such, even though it has been demonstrated that an arbitrary elliptical retarder array can be realized using a cascade of three SLMs \cite{he2023universal} aligned at $0^{\circ}$, $45^{\circ}$, and $0^{\circ}$---the minimum number required for an inverting map to exist---decompositions obtained in this manner are often discontinuous and exhibit the same issues discussed above.

The analysis of continuous parameter decomposition in matter field generation follows a structure similar to that of beam generation, with the main difference that degeneracy now occurs along two circles corresponding to classical gimbal lock rather than at two points. We show in Supplementary Note 2 that, in the case of a four-SLM cascade, the problem reduces to that of the three-SLM case for beam generation, and that obstructions are characterized by a corresponding class in \v{C}ech cohomology.

\begin{figure}[!t]
    \centering
    \includegraphics[width=\textwidth]{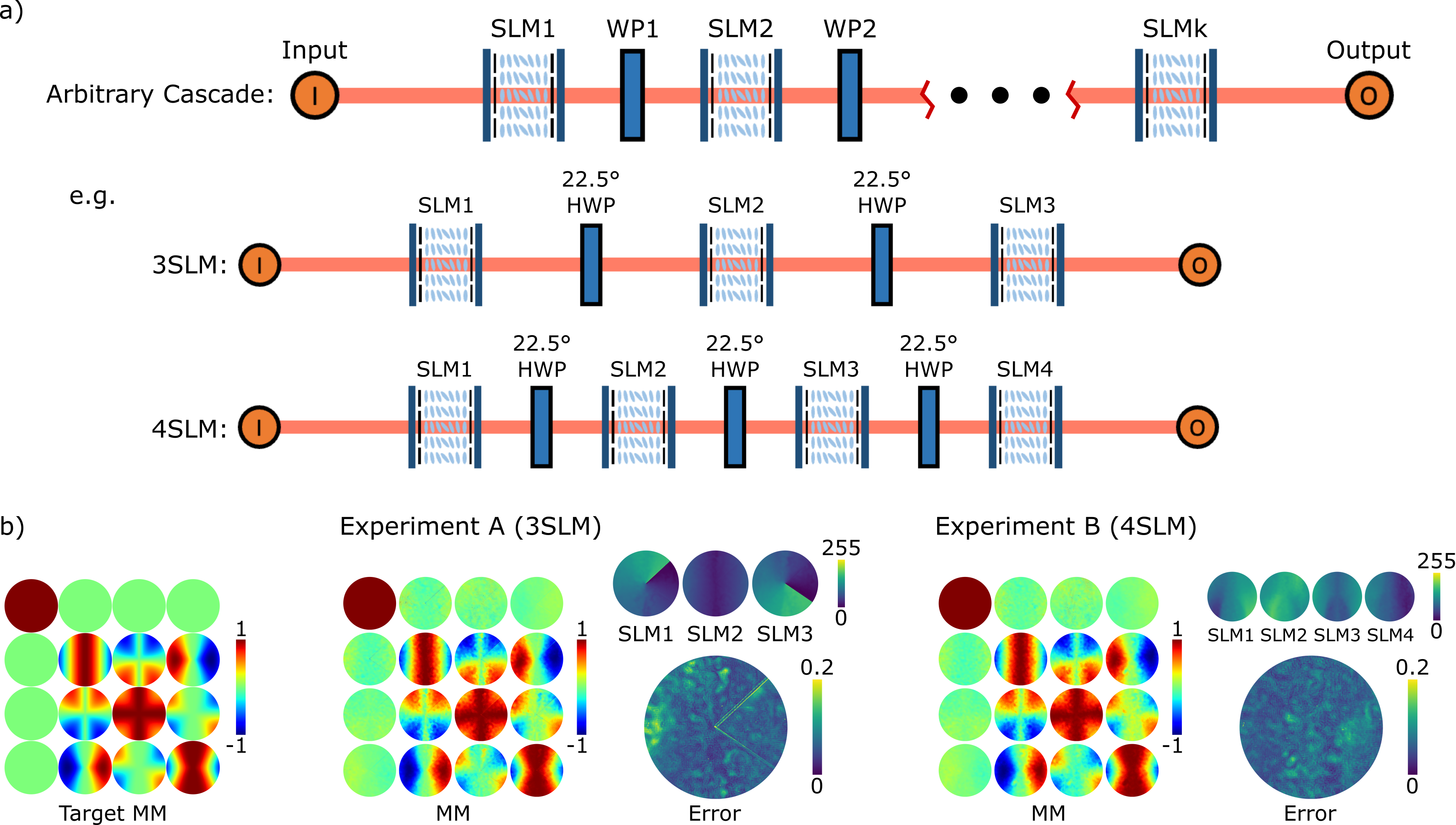}
    \caption{{\bf Experimental results (matter field generation).} a) Experimental configurations adopted in this work. An abstract cascade of arbitrary length $k$ is shown to illustrate the generality of the proposed framework, alongside the concrete three- and four-SLM cascades implemented experimentally. b) Experimentally measured Mueller matrices (MMs) for three- and four-SLM cascades used in matter-field generation. The corresponding phase patterns applied to each SLM and the Frobenius-norm errors between the generated and target fields are also shown. In experiment A, the phase decomposition exhibits discontinuities, resulting in error amplification along the corresponding discontinuity lines.}
\label{fig:matter}
\end{figure}

Fig.\ \ref{fig:matter} presents experimentally measured Mueller matrices for matter field generation using three- and four-SLM cascades, corresponding to experiments A and B, respectively. Note that, in each case, the same cascade structure is used for both beam and matter field generation, as depicted in Fig.\ \ref{fig:matter}a. An abstract cascade of length $k$ is also shown, indicating the applicability of our framework to more complex cascades. For experiment A, the phases in the three-SLM cascade fail to decompose continuously, as the point $r=0$ corresponds to a point of gimbal lock. By contrast, for experiment B, the relevant cohomology class vanishes and continuous decomposition is obtained. The figure also presents normalized Frobenius-norm error distributions of the measured Mueller matrices, which further demonstrate reduced errors when continuous decomposition is achieved, highlighting the importance of this condition for generating high-quality matter fields. A more thorough investigation of matter field generation, including extensions to Jones matrices, is left for future work.

Before concluding, we note several caveats relevant to the present analysis. The first concerns a generic constraint on the number of SLMs required to produce a desired field. Based on dimensional arguments alone, a $k$-dimensional target field requires at least $k$ SLMs for its generation. However, when this minimum number is used, no remaining degrees of freedom are available to enable continuous decomposition. By contrast, introducing one additional SLM reduces continuous parameter decomposition to a lifting problem in which each fiber contains a spatially varying one-dimensional forbidden set. With two additional SLMs, the corresponding forbidden sets are reduced to zero-dimensional sets. This implies, heuristically, that $k+2$ SLMs should always be sufficient to achieve continuous decomposition, although establishing this claim rigorously appears challenging. Next, a practical aspect of this work is its extension to full vectorial control. When intensity, phase, and polarization are simultaneously manipulated, there is no direct topological obstruction to continuous decomposition, since the relevant parameter space is $\mathbb{C}^2$, which is topologically trivial. However, in practice, intensity is typically fixed independently of phase and polarization using cross polarizers, rather than being jointly controlled. In this case, the problem formally reduces to a parameter space $\mathbb{R}\times S^3$, for which $\pi_3(\mathbb{R}\times S^3)=\mathbb{Z}$, so that issues similar to those discussed above may arise. This, in turn, suggests two possible routes for resolving topological obstructions in full vectorial control, namely by treating the phase-polarization problem separately from intensity, or solving all degrees of freedom jointly. 

To conclude, in this paper we discussed the topological obstructions that arise in achieving continuous parameter distributions in two key applications of structured matter, namely beam generation and matter field generation. In both cases, a fundamental mismatch of homotopy groups implies that continuous decomposition, which is of key importance for ensuring the quality of the produced field, can be achieved only for cascades of sufficient length. Moreover, in the context of topological structured light and topological structured matter, continuity is not only essential but necessary to ensure the proper behaviour of topological indices and for topological robustness to be exhibited. As such, we anticipate that the design process introduced in this work will be applicable to future applications involving structured fields, including optical skyrmions \cite{waveguide} and axis-geometry-based skyrmions \cite{chopsticks}, paving the way for broader advances in topological field engineering.

\clearpage

\bibliographystyle{naturemag}
\bibliography{main}

\begin{thebibliography}{10}
\expandafter\ifx\csname url\endcsname\relax
  \def\url#1{\texttt{#1}}\fi
\expandafter\ifx\csname urlprefix\endcsname\relax\def\urlprefix{URL }\fi
\providecommand{\bibinfo}[2]{#2}
\providecommand{\eprint}[2][]{\url{#2}}

\bibitem{He2022}
\bibinfo{author}{He, C.}, \bibinfo{author}{Shen, Y.~J.} \& \bibinfo{author}{Forbes, A.}
\newblock \bibinfo{title}{Towards higher-dimensional structured light}.
\newblock \emph{\bibinfo{journal}{Light: Science \& Applications}} \textbf{\bibinfo{volume}{11}}, \bibinfo{pages}{205} (\bibinfo{year}{2022}).

\bibitem{he_polarisation_2021}
\bibinfo{author}{He, C.} \emph{et~al.}
\newblock \bibinfo{title}{Polarisation optics for biomedical and clinical applications: a review}.
\newblock \emph{\bibinfo{journal}{Light: Science \& Applications}} \textbf{\bibinfo{volume}{10}}, \bibinfo{pages}{194} (\bibinfo{year}{2021}).

\bibitem{Deng2023_H_E_birefringence}
\bibinfo{author}{Deng, L.~Y.} \emph{et~al.}
\newblock \bibinfo{title}{{Influence of hematoxylin and eosin staining on linear birefringence measurement of fibrous tissue structures in polarization microscopy}}.
\newblock \emph{\bibinfo{journal}{Journal of Biomedical Optics}} \textbf{\bibinfo{volume}{28}}, \bibinfo{pages}{102909} (\bibinfo{year}{2023}).

\bibitem{He2023VectorialAdaptiveOptics}
\bibinfo{author}{He, C.}, \bibinfo{author}{Antonello, J.} \& \bibinfo{author}{Booth, M.~J.}
\newblock \bibinfo{title}{Vectorial adaptive optics}.
\newblock \emph{\bibinfo{journal}{eLight}} \textbf{\bibinfo{volume}{3}}, \bibinfo{pages}{23} (\bibinfo{year}{2023}).

\bibitem{He2020_VectorialAdaptiveOptics}
\bibinfo{author}{He, C.}, \bibinfo{author}{Hu, Q.}, \bibinfo{author}{Dai, Y.~Y.} \& \bibinfo{author}{Booth, M.~J.}
\newblock \bibinfo{title}{{Vectorial adaptive optics – correction of polarization and phase}}.
\newblock In \emph{\bibinfo{booktitle}{Imaging and Applied Optics Congress, OSA Technical Digest (Optica Publishing Group)}}, \bibinfo{pages}{paper OF2B.5} (\bibinfo{organization}{Optical Society of America}, \bibinfo{year}{2020}).

\bibitem{Zhang2023_OptLett_6136}
\bibinfo{author}{Zhang, Z.} \emph{et~al.}
\newblock \bibinfo{title}{{Analysis and optimization of aberration induced by oblique incidence for in-vivo tissue polarimetry}}.
\newblock \emph{\bibinfo{journal}{Optics Letters}} \textbf{\bibinfo{volume}{48}}, \bibinfo{pages}{6136--6139} (\bibinfo{year}{2023}).

\bibitem{Shen2022_PolarizationAberrations}
\bibinfo{author}{Shen, Y.~X.} \emph{et~al.}
\newblock \bibinfo{title}{{Polarization Aberrations in High-Numerical-Aperture Lens Systems and Their Effects on Vectorial-Information Sensing}}.
\newblock \emph{\bibinfo{journal}{Remote Sensing}} \textbf{\bibinfo{volume}{14}}, \bibinfo{pages}{1932} (\bibinfo{year}{2022}).

\bibitem{Ma2025OpticalSkyrmions}
\bibinfo{author}{Ma, Y.~F.} \emph{et~al.}
\newblock \bibinfo{title}{Using optical skyrmions to assess vectorial adaptive optics for aberration-aberrated systems}.
\newblock \emph{\bibinfo{journal}{Science Advances}} \textbf{\bibinfo{volume}{11}} (\bibinfo{year}{2025}).

\bibitem{Cheng2025MetrologyWithATwist}
\bibinfo{author}{Cheng, M.~J.} \emph{et~al.}
\newblock \bibinfo{title}{Metrology with a twist: probing and sensing with vortex light}.
\newblock \emph{\bibinfo{journal}{Light: Science \& Applications}} \textbf{\bibinfo{volume}{14}}, \bibinfo{pages}{4} (\bibinfo{year}{2025}).

\bibitem{Willner2021OrbitalAngularMomentumForCommunications}
\bibinfo{author}{Willner, A.~E.} \emph{et~al.}
\newblock \bibinfo{title}{Orbital angular momentum of light for communications}.
\newblock \emph{\bibinfo{journal}{APL Photonics}} \textbf{\bibinfo{volume}{8}}, \bibinfo{pages}{041312} (\bibinfo{year}{2021}).

\bibitem{Wang2025PerturbationResilient}
\bibinfo{author}{Wang, A.~A.} \emph{et~al.}
\newblock \bibinfo{title}{Perturbation-resilient integer arithmetic using optical skyrmions}.
\newblock \emph{\bibinfo{journal}{Nature Photonics}} \textbf{\bibinfo{volume}{19}}, \bibinfo{pages}{1367–1375} (\bibinfo{year}{2025}).

\bibitem{ornelas_non-local_2024}
\bibinfo{author}{Ornelas, P.} \emph{et~al.}
\newblock \bibinfo{title}{Non-local skyrmions as topologically resilient quantum entangled states of light}.
\newblock \emph{\bibinfo{journal}{Nature Photonics}} \textbf{\bibinfo{volume}{18}}, \bibinfo{pages}{258--266} (\bibinfo{year}{2024}).

\bibitem{ornelas2024topologicalrejectionnoisenonlocal}
\bibinfo{author}{Ornelas, P.} \emph{et~al.}
\newblock \bibinfo{title}{Topological rejection of noise by quantum skyrmions}.
\newblock \emph{\bibinfo{journal}{Nature Communications}} \textbf{\bibinfo{volume}{16}}, \bibinfo{pages}{2934} (\bibinfo{year}{2025}).

\bibitem{lu_homogeneous_1994}
\bibinfo{author}{Lu, S.~Y.} \& \bibinfo{author}{Chipman, R.~A.}
\newblock \bibinfo{title}{Homogeneous and inhomogeneous {Jones} matrices}.
\newblock \emph{\bibinfo{journal}{Journal of the Optical Society of America A}} \textbf{\bibinfo{volume}{11}}, \bibinfo{pages}{766--773} (\bibinfo{year}{1994}).

\bibitem{Jorge2022}
\bibinfo{author}{Gil, J.~J.} \emph{et~al.}
\newblock \emph{\bibinfo{title}{Polarized light and the Mueller Matrix Approach. {\normalfont 2nd edn.}}} (\bibinfo{publisher}{Boca Raton: CRC Press}, \bibinfo{year}{2022}).

\bibitem{BornWolf1999PrinciplesOfOptics}
\bibinfo{author}{Born, M.} \& \bibinfo{author}{Wolf, E.}
\newblock \emph{\bibinfo{title}{Principles of Optics. {\normalfont 7th edn.}}} (\bibinfo{publisher}{Cambridge University Press}, \bibinfo{year}{1999}).

\bibitem{Ji2023MetasurfaceDesignQuantumOptics}
\bibinfo{author}{Ji, W.~Y.} \emph{et~al.}
\newblock \bibinfo{title}{Recent advances in metasurface design and quantum optics applications with machine learning, physics-informed neural networks, and topology optimization methods}.
\newblock \emph{\bibinfo{journal}{Light: Science \& Applications}} \textbf{\bibinfo{volume}{12}}, \bibinfo{pages}{169} (\bibinfo{year}{2023}).

\bibitem{yu_light_2011}
\bibinfo{author}{Yu, N.~J.} \emph{et~al.}
\newblock \bibinfo{title}{Light {Propagation} with {Phase} {Discontinuities}: {Generalized} {Laws} of {Reflection} and {Refraction}}.
\newblock \emph{\bibinfo{journal}{Science}} \textbf{\bibinfo{volume}{334}}, \bibinfo{pages}{333--337} (\bibinfo{year}{2011}).

\bibitem{balthasar_mueller_metasurface_2017}
\bibinfo{author}{Balthasar~Mueller, J.~P.} \emph{et~al.}
\newblock \bibinfo{title}{Metasurface {Polarization} {Optics}: {Independent} {Phase} {Control} of {Arbitrary} {Orthogonal} {States} of {Polarization}}.
\newblock \emph{\bibinfo{journal}{Physical Review Letters}} \textbf{\bibinfo{volume}{118}}, \bibinfo{pages}{113901} (\bibinfo{year}{2017}).

\bibitem{yu_broadband_2012}
\bibinfo{author}{Yu, N.~F.} \emph{et~al.}
\newblock \bibinfo{title}{A {Broadband}, {Background}-{Free} {Quarter}-{Wave} {Plate} {Based} on {Plasmonic} {Metasurfaces}}.
\newblock \emph{\bibinfo{journal}{Nano Letters}} \textbf{\bibinfo{volume}{12}}, \bibinfo{pages}{6328--6333} (\bibinfo{year}{2012}).

\bibitem{Chen2012DualPolarityPlasmonicMetalens}
\bibinfo{author}{Chen, X.~Z.} \emph{et~al.}
\newblock \bibinfo{title}{Dual-polarity plasmonic metalens for visible light}.
\newblock \emph{\bibinfo{journal}{Nature Communications}} \textbf{\bibinfo{volume}{3}}, \bibinfo{pages}{1198} (\bibinfo{year}{2012}).

\bibitem{Yang2023_LC_SLM_Review}
\bibinfo{author}{Yang, Y.~Q.}, \bibinfo{author}{Forbes, A.} \& \bibinfo{author}{Cao, L.~C.}
\newblock \bibinfo{title}{{A review of liquid crystal spatial light modulators: devices and applications}}.
\newblock \emph{\bibinfo{journal}{Opto-Electronic Science}} \textbf{\bibinfo{volume}{2}}, \bibinfo{pages}{230026} (\bibinfo{year}{2023}).

\bibitem{Converter}
\bibinfo{author}{Zhang, R.~C.} \emph{et~al.}
\newblock \bibinfo{title}{Multiplexed vector beam conversion via complex structured matter}.
\newblock \bibinfo{howpublished}{Preprint at \url{https://arxiv.org/abs/2512.22980}} (\bibinfo{year}{2025}).

\bibitem{Zhang2025Elliptical}
\bibinfo{author}{Zhang, R.~C.} \emph{et~al.}
\newblock \bibinfo{title}{Elliptical vectorial metrics for physically plausible polarization information analysis}.
\newblock \emph{\bibinfo{journal}{Advanced Photonics Nexus}} \textbf{\bibinfo{volume}{4}}, \bibinfo{pages}{066015} (\bibinfo{year}{2025}).

\bibitem{he2023universal}
\bibinfo{author}{He, C.} \emph{et~al.}
\newblock \bibinfo{title}{A reconfigurable arbitrary retarder array as complex structured matter}.
\newblock \emph{\bibinfo{journal}{Nature Communications}} \textbf{\bibinfo{volume}{16}}, \bibinfo{pages}{4902} (\bibinfo{year}{2025}).

\bibitem{wang2024topological}
\bibinfo{author}{Wang, A.~A.} \emph{et~al.}
\newblock \bibinfo{title}{Topological protection of optical skyrmions through complex media}.
\newblock \emph{\bibinfo{journal}{Light: Science \& Applications}} \textbf{\bibinfo{volume}{13}}, \bibinfo{pages}{314} (\bibinfo{year}{2024}).

\bibitem{Han2013VectorialOpticalFieldGenerator}
\bibinfo{author}{Han, W.} \emph{et~al.}
\newblock \bibinfo{title}{Vectorial optical field generator for the creation of arbitrarily complex fields}.
\newblock \emph{\bibinfo{journal}{Optics Express}} \textbf{\bibinfo{volume}{21}}, \bibinfo{pages}{20692--20706} (\bibinfo{year}{2013}).

\bibitem{Rong2014GenerationArbitraryVectorBeams}
\bibinfo{author}{Rong, Z.~Y.} \emph{et~al.}
\newblock \bibinfo{title}{Generation of arbitrary vector beams with cascaded liquid-crystal polarization converters}.
\newblock \emph{\bibinfo{journal}{Optics Express}} \textbf{\bibinfo{volume}{22}}, \bibinfo{pages}{1636--1644} (\bibinfo{year}{2014}).

\bibitem{chen2023superresolution}
\bibinfo{author}{Chen, X.} \emph{et~al.}
\newblock \bibinfo{title}{Superresolution structured illumination microscopy reconstruction algorithms: a review}.
\newblock \emph{\bibinfo{journal}{Light: Science \& Applications}} \textbf{\bibinfo{volume}{12}}, \bibinfo{pages}{172} (\bibinfo{year}{2023}).

\bibitem{shen_optical_2023}
\bibinfo{author}{Shen, Y.~J.} \emph{et~al.}
\newblock \bibinfo{title}{Optical skyrmions and other topological quasiparticles of light}.
\newblock \emph{\bibinfo{journal}{Nature Photonics}} \textbf{\bibinfo{volume}{18}}, \bibinfo{pages}{15--25} (\bibinfo{year}{2024}).

\bibitem{Tsesses2018}
\bibinfo{author}{Tsesses, S.} \emph{et~al.}
\newblock \bibinfo{title}{Optical skyrmion lattice in evanescent electromagnetic fields}.
\newblock \emph{\bibinfo{journal}{Science}} \textbf{\bibinfo{volume}{361}}, \bibinfo{pages}{993–996} (\bibinfo{year}{2018}).

\bibitem{Gao2020}
\bibinfo{author}{Gao, S.~J.} \emph{et~al.}
\newblock \bibinfo{title}{Paraxial skyrmionic beams}.
\newblock \emph{\bibinfo{journal}{Physical Review A}} \textbf{\bibinfo{volume}{102}}, \bibinfo{pages}{053513} (\bibinfo{year}{2020}).

\bibitem{Shen2023}
\bibinfo{author}{Shen, Y.~J.} \emph{et~al.}
\newblock \bibinfo{title}{Topologically controlled multiskyrmions in photonic gradient-index lenses}.
\newblock \emph{\bibinfo{journal}{Physical Review Applied}} \textbf{\bibinfo{volume}{21}}, \bibinfo{pages}{024025} (\bibinfo{year}{2024}).

\bibitem{Beckley2010FullPoincareBeams}
\bibinfo{author}{Beckley, A.~M.}, \bibinfo{author}{Brown, T.~G.} \& \bibinfo{author}{Alonso, M.~A.}
\newblock \bibinfo{title}{{F}ull {P}oincar\'e beams}.
\newblock \emph{\bibinfo{journal}{Optics Express}} \textbf{\bibinfo{volume}{18}}, \bibinfo{pages}{10777--10785} (\bibinfo{year}{2010}).

\bibitem{Azzam16}
\bibinfo{author}{Azzam, R. M.~A.}
\newblock \bibinfo{title}{Stokes-vector and {M}ueller-matrix polarimetry [{I}nvited]}.
\newblock \emph{\bibinfo{journal}{Journal of the Optical Society of America A}} \textbf{\bibinfo{volume}{33}}, \bibinfo{pages}{1396--1408} (\bibinfo{year}{2016}).

\bibitem{Azzam78}
\bibinfo{author}{Azzam, R. M.~A.}
\newblock \bibinfo{title}{{P}hotopolarimetric measurement of the {M}ueller matrix by {F}ourier analysis of a single detected signal}.
\newblock \emph{\bibinfo{journal}{Optics Letters}} \textbf{\bibinfo{volume}{2}}, \bibinfo{pages}{148--150} (\bibinfo{year}{1978}).

\bibitem{Yuxi}
\bibinfo{author}{Cai, Y.~X.} \emph{et~al.}
\newblock \bibinfo{title}{Rethinking conditioning in polarimetry: a new framework beyond $\ell^2$-based metrics}.
\newblock \bibinfo{howpublished}{Preprint at \url{https://arxiv.org/abs/2508.16483}} (\bibinfo{year}{2025}).

\bibitem{Zeng2025UltraHighOrderSkyrmions}
\bibinfo{author}{Zeng, X.~J.} \emph{et~al.}
\newblock \bibinfo{title}{{Tailoring Ultra-High-Order Optical Skyrmions}}.
\newblock \emph{\bibinfo{journal}{Laser \& Photonics Reviews}} \textbf{\bibinfo{volume}{19}} (\bibinfo{year}{2025}).

\bibitem{Shen2022OpticalSkyrmions}
\bibinfo{author}{Shen, Y.~J.} \emph{et~al.}
\newblock \bibinfo{title}{{Generation of Optical Skyrmions with Tunable Topological Textures}}.
\newblock \emph{\bibinfo{journal}{ACS Photonics}} \textbf{\bibinfo{volume}{9}}, \bibinfo{pages}{296--303} (\bibinfo{year}{2022}).

\bibitem{Liu2022}
\bibinfo{author}{Liu, C.~X.} \emph{et~al.}
\newblock \bibinfo{title}{Disorder-induced topological state transition in the optical skyrmion family}.
\newblock \emph{\bibinfo{journal}{Physical Review Letters}} \textbf{\bibinfo{volume}{129}}, \bibinfo{pages}{267401} (\bibinfo{year}{2022}).

\bibitem{wang2024generalizedskyrmions}
\bibinfo{author}{Wang, A.~A.} \emph{et~al.}
\newblock \bibinfo{title}{Generalized skyrmions}.
\newblock \bibinfo{howpublished}{Preprint at \url{https://arxiv.org/abs/2409.17390}} (\bibinfo{year}{2024}).

\bibitem{Hu2021ArbitraryComplexRetarders}
\bibinfo{author}{Hu, Q.}, \bibinfo{author}{He, C.} \& \bibinfo{author}{Booth, M.~J.}
\newblock \bibinfo{title}{Arbitrary complex retarders using a sequence of spatial light modulators as the basis for adaptive polarisation compensation}.
\newblock \emph{\bibinfo{journal}{Journal of Optics}} \textbf{\bibinfo{volume}{23}}, \bibinfo{pages}{065602} (\bibinfo{year}{2021}).

\bibitem{Ma2024Vectorial}
\bibinfo{author}{Ma, Y.~F.} \emph{et~al.}
\newblock \bibinfo{title}{Vectorial adaptive optics for advanced imaging systems}.
\newblock \emph{\bibinfo{journal}{Journal of Optics}} \textbf{\bibinfo{volume}{26}}, \bibinfo{pages}{065402} (\bibinfo{year}{2024}).

\bibitem{chopsticks}
\bibinfo{author}{Zhang, Y.~Q.} \emph{et~al.}
\newblock \bibinfo{title}{Skyrmions based on optical anisotropy for topological encoding}.
\newblock \bibinfo{howpublished}{Preprint at \url{https://arxiv.org/abs/2508.16483}} (\bibinfo{year}{2025}).

\bibitem{hu_arbitrary_2020}
\bibinfo{author}{Hu, Q.} \emph{et~al.}
\newblock \bibinfo{title}{Arbitrary vectorial state conversion using liquid crystal spatial light modulators}.
\newblock \emph{\bibinfo{journal}{Optics Communications}} \textbf{\bibinfo{volume}{459}}, \bibinfo{pages}{125028} (\bibinfo{year}{2020}).

\bibitem{waveguide}
\bibinfo{author}{Wang, A.~A.} \emph{et~al.}
\newblock \bibinfo{title}{{O}ptical {S}kyrmions in {W}aveguides}.
\newblock \bibinfo{howpublished}{Preprint at \url{https://arxiv.org/abs/2505.06735}} (\bibinfo{year}{2025}).

\end{thebibliography}


\begin{thebibliography}{1}
\expandafter\ifx\csname url\endcsname\relax
  \def\url#1{\texttt{#1}}\fi
\expandafter\ifx\csname urlprefix\endcsname\relax\def\urlprefix{URL }\fi
\providecommand{\bibinfo}[2]{#2}
\providecommand{\eprint}[2][]{\url{#2}}

\bibitem{Liu2020DielectricResonanceMetasurfaces}
\bibinfo{author}{Liu, W.~W.} \emph{et~al.}
\newblock \bibinfo{title}{Dielectric resonance-based optical metasurfaces: From fundamentals to applications}.
\newblock \emph{\bibinfo{journal}{iScience}} \textbf{\bibinfo{volume}{23}}, \bibinfo{pages}{101868} (\bibinfo{year}{2020}).

\bibitem{Babicheva2024MieResonantMetaphotonics}
\bibinfo{author}{Babicheva, V.~E.} \& \bibinfo{author}{Evlyukhin, A.~B.}
\newblock \bibinfo{title}{Mie-resonant metaphotonics}.
\newblock \emph{\bibinfo{journal}{Advances in Optics and Photonics}} \textbf{\bibinfo{volume}{16}}, \bibinfo{pages}{539--658} (\bibinfo{year}{2024}).

\bibitem{Overvig2019DielectricMetasurfaces}
\bibinfo{author}{Overvig, A.~C.} \emph{et~al.}
\newblock \bibinfo{title}{Dielectric metasurfaces for complete and independent control of the optical amplitude and phase}.
\newblock \emph{\bibinfo{journal}{Light: Science \& Applications}} \textbf{\bibinfo{volume}{8}}, \bibinfo{pages}{92} (\bibinfo{year}{2019}).

\bibitem{Kim2025AntiAliasedMetasurfaces}
\bibinfo{author}{Kim, S.} \emph{et~al.}
\newblock \bibinfo{title}{Anti-aliased metasurfaces beyond the nyquist limit}.
\newblock \emph{\bibinfo{journal}{Nature Communications}} \textbf{\bibinfo{volume}{16}}, \bibinfo{pages}{411} (\bibinfo{year}{2025}).

\end{thebibliography}

\clearpage

\section*{Methods}
\setcounter{section}{0}
\section{Importance of continuous decomposition of parameters}

\begin{figure}[!t]
    \centering
    \includegraphics[width=\textwidth]{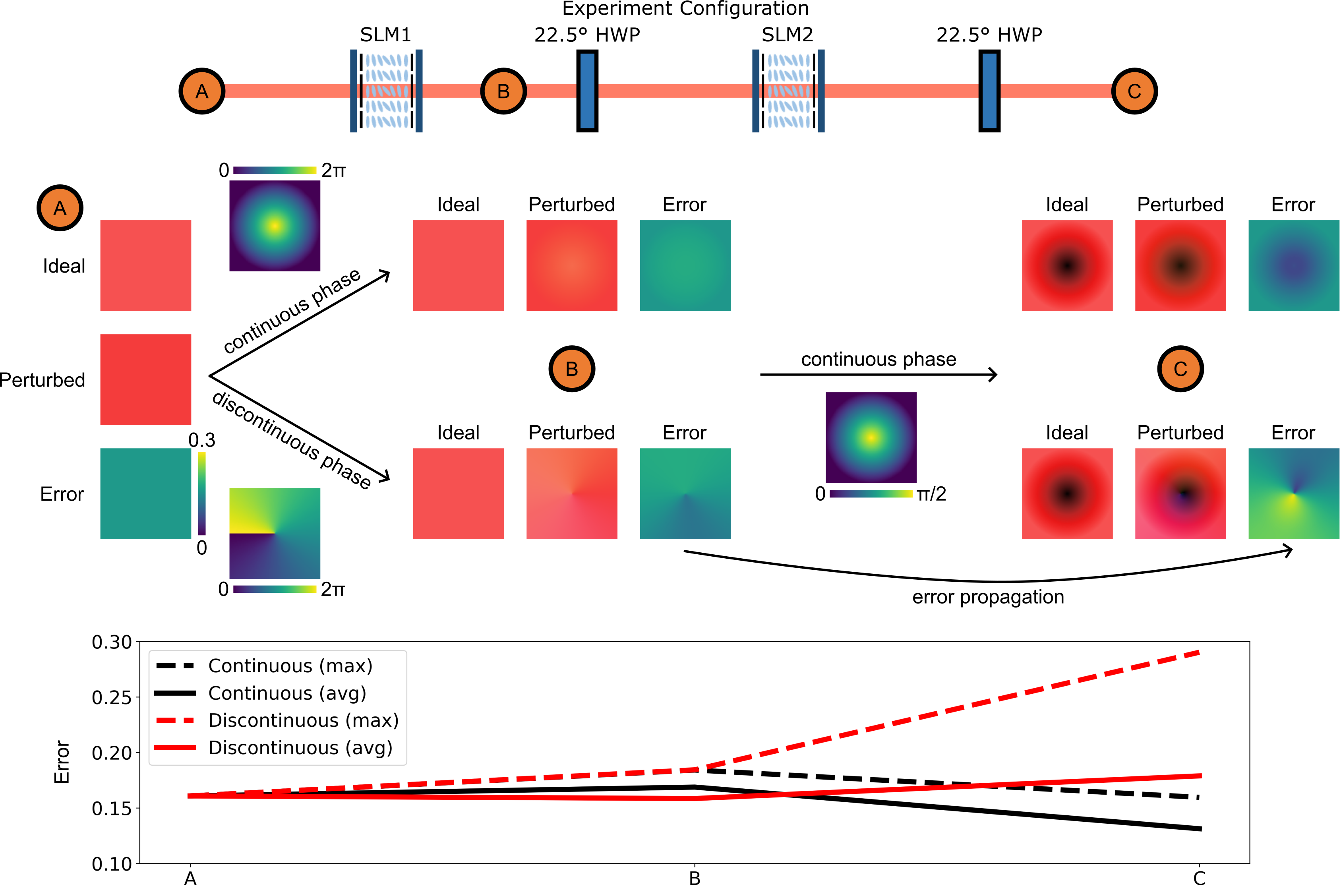}
    \caption{{\bf Error Propagation.} Ideal and perturbed Stokes fields propagating through a two-SLM cascade beam generator, with the corresponding experimental configuration shown at the top of the figure. Fields at three stages are provided: immediately before the first SLM, immediately after the first SLM, and at the system output. The spatial distribution of the $\ell^2$ error between the ideal and perturbed states at each stage is also shown. Finally, a line plot of the mean and maximum $\ell^2$ errors is presented.}
\label{fig:methods}
\end{figure}

As discussed in the main text, discontinuities in the cascade parameter distribution can lead to detrimental engineering consequences. To elaborate, discontinuities that arise due to topological obstructions typically reflect underlying degeneracies (Supplementary Note 1); for example, any rotation about the $s_1$-axis leaves horizontally or vertically polarized light unchanged, even though rotations by different angles about the $s_1$-axis are represented by very different matrices. Therefore, if horizontally polarized light is incident on a linear retarder with its axis aligned at $0^\circ$, any distribution of retardance will leave the polarization state unchanged, and the transformation appears ``continuous'' even if the underlying matter field is not. In such a scenario, if the incident light is not perfectly horizontally or vertically polarized, discontinuities in the underlying retardance field lead to unintended and effectively uncontrollable polarization transformations at different points in space, thereby amplifying existing errors that compound in a cascaded structure and can be difficult to manage (Fig.\ \ref{fig:methods}). 

As such, when parameter decomposition is discontinuous, fabrication limitations such as parameter quantization, material nonuniformity, and finite spatial resolution, along with experimental imperfections such as misalignments, environmental sensitivity, and angular errors, can strongly affect system performance near points of discontinuity. Practical examples of such limitations in liquid crystal devices include the temperature dependence of the device’s retardance, pixel-to-pixel non-uniformities, discretization of voltage levels, and non-idealities such as absorption and scattering that give rise to diattenuation and depolarization. In metasurface-based optical retarders, this problem is even more pronounced, as subwavelength feature sizes introduce fabrication challenges that restrict the available library of meta-atoms, thereby limiting the range of achievable parameters, while simultaneously imposing constraints on the fidelity of the synthesized metasurface \citesupp{Liu2020DielectricResonanceMetasurfaces}. For example, in both Mie-type resonance-based metasurfaces \citesupp{Babicheva2024MieResonantMetaphotonics} and dielectric nanopillar metasurfaces \citesupp{Overvig2019DielectricMetasurfaces}, the optical response is governed by the geometry of the individual meta-atoms and is therefore limited by lithographic resolution, whereas in PB-phase metasurfaces \cite{Chen2012DualPolarityPlasmonicMetalens} the meta-atoms are geometrically identical and control is instead limited by the discretization of their in-plane orientation. 

We emphasize that the fabrication limitations discussed above persist even when the parameter decomposition is continuous; however, it is the increased sensitivity of system performance to errors in the presence of discontinuities that magnifies their impact. A separate but important consideration is that, at points of discontinuity, the desired parameter distribution varies rapidly in space and therefore cannot be practically realized due to the finite size of meta-atoms \citesupp{Kim2025AntiAliasedMetasurfaces}. Thus, discontinuous decompositions not only exacerbate error propagation throughout the cascade, but also increase the deviation between the true optical response of each individual layer and its designed parameters.

Another consequence of a parameter distribution that varies rapidly in space is that the geometry of the meta-atoms becomes highly non-uniform, which can introduce additional fabrication challenges and unwanted aberrations arising from edge effects. For example, abrupt variations in the height of dielectric nanopillars necessarily lead to increased scattering, resulting in both intensity and phase aberrations that strongly degrade the quality of the generated field.

Lastly, we present theoretical results validating the amplification of errors arising from discontinuous parameter distributions in Fig.\ \ref{fig:methods}. In this simulation, a two-SLM cascade beam generator is considered, with the designed polarization field labeled \textbf{Ideal} in the figure. Two different decompositions of the same transformation are considered, differing only in the phase distribution applied to the first SLM---continuous in one case and discontinuous in the other---while the same continuous phase is applied to the second SLM in both cases. The figure also illustrates the system under perturbation. Perturbations are introduced in two ways: first, the incident field is slightly rotated relative to the designed polarization state; second, a small, randomly generated angular error in the range $[-2.0^\circ,2.0^\circ]$ and a randomly generated depolarization in the range $[0.95,1.0]$ are applied to each SLM in the cascade. Lastly, the $\ell^2$-error distribution at each stage of the cascade is shown, together with the corresponding average and maximum values indicated in the line plot.

From Fig.\ \ref{fig:methods}, the amplification of error due to discontinuous parameters is clearly evident. After the first SLM, the error distributions in the continuous and discontinuous cases are comparable. However, after passing through the second SLM, the maximum error in the discontinuous case increases to more than twice that of the continuous case, highlighting the detrimental impact of enhanced error sensitivity that motivates this work. 

\section{Topological obstructions in the three-SLM case}

In this section, we present a concrete example demonstrating the failure of continuous phase decomposition in the three-SLM case and highlighting the role of the topology of $\Sigma$.

If $\Sigma \cong S^1$, then $\tau \coloneqq s_2\lvert_\Sigma+\mathrm{i}s_3\lvert_\Sigma$ defines a map $\tau \colon S^1 \longrightarrow S^1$. Suppose that there exists a continuous $\psi_1 \lvert_\Sigma \colon S^1 \longrightarrow S^1$ satisfying $\psi_1\lvert_\Sigma(\theta) \neq \tau(\theta)$ for every $\theta \in S^1$. It follows that $\psi_1\lvert_\Sigma(\theta) \bar{\tau}(\theta) \neq 1$ so the map $\psi_1\lvert_\Sigma \bar{\tau} \colon S^1 \longrightarrow S^1$ fails to be surjective and therefore has degree 0. Consequently, $\deg \psi_1\lvert_\Sigma = \deg \tau$. Thus, whenever $\deg \tau \neq 0$, a globally defined continuous $\psi_1$ satisfying the required conditions cannot exist.

\bibliographystylesupp{naturemag}
{\bibliographysupp{main}}

\end{document}


\maketitle

\section{Continuous decomposition in beam generation}
\subsection{The 3 SLM case}
As discussed in the main text, the key obstruction to the existence of a globally defined and continuous $\psi_1$ is whether $\tau \colon \Sigma \longrightarrow  S^1$ can be lifted to $\mathbb{R}$. To see why this is necessary, suppose there exists a $\hat{\psi}_1 \colon \mathbb{R}^2 \longrightarrow \mathbb{R}$ such that $e^{\mathrm{i}\hat{\psi}_1(z)} = \psi_1(z) \neq \tau (z)$ for every $z \in \Sigma$. Let $\arg$ denote a continuous branch of the argument on $S^1 \backslash \{1\}$. Since $\psi_1(z)\,\bar{\tau}(z) \neq 1$ for all $z \in \Sigma$, the composition $\arg(\psi_1\bar{\tau})$ is well defined and continuous on $\Sigma$. Thus, 
\begin{equation*}
    \hat{\tau} = \hat{\psi}_1 - \arg(\psi_1\bar{\tau})
\end{equation*}
is a lift of $\tau$ to $\mathbb{R}$. Conversely, suppose $\tau \colon \Sigma \longrightarrow S^1$ can be lifted to $\hat{\tau} \colon \Sigma \longrightarrow \mathbb{R}$, then $\hat{\psi}_1\lvert_\Sigma = \hat{\tau}+\pi/4$ satisfies $e^{\mathrm{i}\hat{\psi}_1(z)} \neq \pm \tau(z)$ for every $z \in \Sigma$. By the Tietze extension theorem, $\hat{\psi}_1|_\Sigma$ can always be extended to all of $\mathbb{R}^2$, and the extension satisfies all the required properties.

Therefore, the problem of finding a continuous decomposition reduces to determining whether the map $\tau \colon \Sigma \longrightarrow S^1$ admits a lift to $\mathbb{R}$. To address this in general, note that the map $\tau \colon \Sigma \longrightarrow S^1$ can be lifted locally in the following way. Let $z \in \Sigma$ and choose any $q\in \mathbb{R}$ such that $e^{\mathrm{i}q} = \tau(z)$. The set $U = \{e^{\mathrm{i}p} \colon q-\pi/4<p<q+\pi/4\}$ is clearly open in $S^1$ so $\tau^{-1}(U)$ is an open neighbourhood of $z$ in $\Sigma$. By selecting the trivial lift on $\tau^{-1}(U)$ taking values in $(q - \pi/4,q + \pi/4)$, we obtain a local lift of $\tau$ around $z$.  
Thus, for every $z \in \Sigma$ there exists an open neighbourhood on which $\tau$ admits a lift. This is a concrete rewriting of the more commonly encountered 
sheaf-theoretic formulation, namely the surjectivity of 
$\exp(\mathrm{i}\,\cdot)$ in the short exact sequence
\begin{equation*}
    0 \rightarrow \underline{\mathbb{Z}} \hookrightarrow \underline{\mathbb{R}} \xrightarrow{\exp(\mathrm{i}\cdot)} \underline{S}^1\rightarrow 0. 
\end{equation*}

Now, let $\mathfrak{U} = \{U_i\}_{i\in I}$ be an open cover of $\Sigma$, and let $\hat{\tau}_i\colon U_i \longrightarrow \mathbb{R}$ be lifts of $\tau$ on each $U_i$. We first recall the definition of the \v{C}ech cochain complex $C^\bullet(\mathfrak{U}; \mathbb{Z})$ associated with the cover $\mathfrak{U}$. For each $k \geq 0$, the group of $k$-cochains is
\begin{equation*}
    C^k(\mathfrak{U}; \mathbb{Z}) = \prod_{i_0,\dots,i_k \in I} C^0(U_{i_0} \cap \cdots \cap U_{i_k}; \mathbb{Z}) = \{\, c_{i_0 \dots i_k} \colon U_{i_0} \cap \cdots \cap U_{i_k} \longrightarrow \mathbb{Z} \,\},
\end{equation*}
where $c_{i_0 \dots i_k}$ is a locally constant integer-valued function on the 
$(k+1)$-fold intersection, and the \v{C}ech differential 
$\delta \colon C^k(\mathfrak{U}; \mathbb{Z}) \longrightarrow C^{k+1}(\mathfrak{U}; \mathbb{Z})$ 
is defined by
\begin{equation*}
    (\delta c)_{i_0 \dots i_{k+1}}
    = \sum_{j=0}^{k+1} (-1)^j
      c_{i_0 \dots \widehat{i_j} \dots i_{k+1}}
      \Big|_{\,U_{i_0} \cap \cdots \cap U_{i_{k+1}}},
\end{equation*}
where the hat denotes omission of the index $i_j$.

If $U_i \cap U_j \neq \varnothing$, since $\tau_i = e^{\mathrm{i}\hat{\tau}_i}$ agrees with $\tau_j = e^{\mathrm{i}\hat{\tau}_j}$ on $U_i\cap U_j$, one has that
\begin{equation*}
    n_{ij} \coloneqq \hat{\tau}_i\lvert_{U_i\cap U_j} -\hat{\tau}_j\lvert_{U_i \cap U_j} \colon U_i\cap U_j \longrightarrow \mathbb{Z}
\end{equation*}
and hence defines a cochain $n \in C^1(\mathfrak{U};\mathbb{Z})$. Moreover, the functions $n_{ij}$ trivially satisfy the cocycle condition $n_{ij} + n_{jk} + n_{ki} = 0$ on $U_i \cap U_j \cap U_k$, which is precisely the statement that $\delta n = 0$. Suppose we now ask whether it is possible to piece together the $\hat{\tau}_i$ into a global function; it is easy to see that this is the case precisely when there exist functions $k_i \colon U_i \longrightarrow \mathbb{Z}$ such that $\hat{\tau}_i - k_i = \hat{\tau}_j - k_j$ on $U_i \cap U_j$. Equivalently, one has 
\begin{equation*}
    n_{ij} = k_i - k_j,
\end{equation*}
which is just the statement that $n = \delta k$ is a coboundary. Taken together, we can rewrite the condition for lifting as the statement
\begin{equation*}
    \text{$\tau$ can be lifted if and only if $[n] = 0 \in H^1(\mathfrak{U}; \mathbb{Z})$}.
\end{equation*}

Passing from the open cover $\mathfrak{U}$ to the direct limit 
\begin{equation*}
    \check{H}^k(\Sigma;\mathbb{Z})\coloneqq \varinjlim_{\mathfrak{V} \succ \mathfrak{U}}\check{H}^k(\mathfrak{V};\mathbb{Z}),
\end{equation*}
we note that $[n]$ determines a cohomology class in $\check{H}^1(\Sigma;\mathbb{Z})$ independent of the initial
cover $\mathfrak{U}$, which we denote by $[\tau]$. It can then be shown that the lifting condition can be expressed equivalently as
\begin{equation*}
    \text{$\tau$ can be lifted if and only if $[\tau] = 0 \in \check{H}^1(\Sigma;\mathbb{Z})$},
\end{equation*}
reducing the entire problem to computing a cohomology class. Note that two important simplifications can be made if $\Sigma$ is sufficiently regular, which allow for a swift computation of $[\tau]$. The first is the case in which $\Sigma$ is locally path-connected, so that standard 
covering space theory reduces the lifting problem to computing the induced map 
$\tau_\ast$ on the fundamental group. 
If, on the other hand, $\Sigma$ is locally contractible, then 
$\check{H}^1(\Sigma;\mathbb{Z})$ is naturally isomorphic to the singular 
cohomology group $H^1(\Sigma;\mathbb{Z})$, and under this identification 
the class $[\tau]$ corresponds to the pullback $\tau^\ast \eta$ where $\eta$ is the canonical generator of $H^1(S^1;\mathbb{Z})$. The latter case holds in most situations, such as in skyrmion generation, and is generally the cleanest approach for determining whether obstructions to lifting exist.

\subsection{The 4 SLM case}

As in the three-SLM case, the problem of finding a field $\tilde{\mathcal{S}}$ such that 
\begin{equation*}
    \tilde{\mathcal{S}}(x,y) \neq \left\{ \pm \mathcal{S}(x,y),\, \pm(1,0,0) \right\}\quad \text{for all } (x,y)\in\mathbb{R}^2
\end{equation*}
can be phrased as a lifting problem, only now to the subspace $E \subset \mathbb{R}^2 \times S^2$ obtained by removing the forbidden points above each $(x,y)\in\mathbb{R}^2$. Moreover, it is clear that this lifting problem is easier than the corresponding one for three SLMs, where the analog of $\tilde{\mathcal{S}}$ is constrained to lie on an $S^1$ submanifold of $S^2$, rather than on all of $S^2$ minus up to four points.
However, the topological obstructions that arise in this case are more difficult to analyze because $E$ is not a genuine fiber bundle when there exist points $(x,y)$ such that $\mathcal{S}(x,y) = \pm(1,0,0)$.  
Note that if this does not occur, then $E$ is an $(S^{2}\backslash \{\text{4 points}\})$-fiber bundle over $\mathbb{R}^2$, which is automatically trivial since $\mathbb{R}^2$ is contractible. Thus, in this case there are no obstructions to the lifting problem, and a continuous decomposition of phases can be found.

More generally, we note that there are no hairy-ball obstructions to finding a field $\tilde{\mathcal{S}}$ with $\tilde{\mathcal{S}}(x,y) \neq \pm \mathcal{S}(x,y)$, since the contractibility of $\mathbb{R}^2$ implies that the pullback of the unit tangent bundle $U(TS^2)$ along $\mathcal{S}$ is trivial. Thus we can always choose a section $\mathcal{T}$ of $\mathcal{S}^\ast U(TS^2)$, from which it follows that
\begin{equation*}
    \exp_{\mathcal{S}(x,y)}(\delta(x,y)\mathcal{T}(x,y))
\end{equation*}
is continuous for every continuous function $\delta$, and is not equal to $\pm \mathcal{S}(x,y)$ provided that $\delta(x,y) \notin \pi\mathbb{Z}$. The difficulty lies in ensuring that the above does not hit $\pm(1,0,0)$, and we conjecture that this is, in general, not possible. Nonetheless, we demonstrate experimentally that, for skyrmions, by adjusting the phases $\hat{\psi}_1$ and $\hat{\psi}_2$, it is possible to push any phase discontinuities to the boundary of the field. We leave a more detailed mathematical analysis of this problem as a possible extension of this work.

\section{Matter field generation}

Consider a three SLM cascade of aligned at $0^{\circ}$, $45^{\circ}$, and $0^{\circ}$. If $\hat{\chi}$, $\hat{\phi}$ and $\hat{\varphi}$ denote the phases of each SLM, respectively, their combined Mueller matrix is given by 
\begin{equation*}
    \begin{pmatrix}
        \cos\hat{\phi} & \sin\hat{\phi}\sin\hat{\chi} & \sin\hat{\phi}\cos\hat{\chi} \\ \sin\hat{\varphi}\sin\hat{\phi} & -\sin\hat{\varphi}\cos\hat{\phi}\sin\hat{\chi}+\cos\hat{\varphi}\cos\hat{\chi} & -\cos\hat{\varphi}\sin\hat{\chi} - \sin\hat{\varphi}\cos\hat{\phi}\cos\hat{\chi} \\ -\cos\hat{\varphi}\sin\hat{\phi} & \cos\hat{\varphi}\cos\hat{\phi}\sin\hat{\chi}+\sin\hat{\varphi}\cos\hat{\chi} & -\sin\hat{\varphi}\sin\hat{\chi}+\cos\hat{\varphi}\cos\hat{\phi}\cos\hat{\chi}
    \end{pmatrix}.
\end{equation*}
Thus, if $\sin \hat{\phi} \neq 0$, there are two continuously varying branches
\begin{align*}
    \hat{\phi} & = \pm \arccos(M_{11}), \\
    \hat{\chi} & = \operatorname{atan2}(M_{12}/\!\sin\hat{\phi}, M_{13}/\!\sin\hat{\phi}), \\
    \hat{\varphi} &= \operatorname{atan2}(M_{21}/\!\sin\hat{\phi}, -M_{31}/\!\sin\hat{\phi}),
\end{align*}
which produce the target Mueller matrix. However, as in the case of beam generation, as $\sin \hat{\phi} \rightarrow 0$, there is no guarantee that $\hat{\chi}$ and $\hat{\varphi}$ approach the same value from all directions. Thus, for a continuous decomposition of parameters, a sufficient condition is to avoid the set given by $\sin \hat{\phi} = 0$. This corresponds to Mueller matrices which take the form
\begin{equation*}
    \begin{pmatrix}
        1 & 0 & 0 \\ 0 & \cos(\hat{\chi}+\hat{\varphi}) & -\sin(\hat{\chi}+\hat{\varphi}) \\ 0 & \sin(\hat{\chi}+\hat{\varphi}) & \cos(\hat{\chi}+\hat{\varphi})
    \end{pmatrix}  \quad \text{and} \quad  \begin{pmatrix}
        -1 & 0 & 0 \\ 0 & \cos(\hat{\chi}-\hat{\varphi}) & -\sin(\hat{\chi}-\hat{\varphi})\\ 0 & -\sin(\hat{\chi}-\hat{\varphi}) & - \cos(\hat{\chi}-\hat{\varphi})
    \end{pmatrix},
\end{equation*}
which arise in the classical problem of gimbal lock. As in the strategy presented in Supplementary Note 1, adding additional SLMs offers additional degrees of freedom that can be used to avoid these problematic points, thus allowing for a continuous decomposition of parameters. 

For example, consider the four-SLM beam generation setup presented in the main text, in which a half-wave plate with its fast axis aligned at $22.5^\circ$ is followed by an SLM with phase $\hat{\psi}_1$ and its fast axis aligned at $0^\circ$. In this case, gimbal lock can be avoided provided if $\hat{\psi}_1$ can be picked such that
\begin{equation*}
    M_{21}\cos\hat{\psi}_1+M_{31}\sin\hat{\psi}_1 \neq \pm 1
\end{equation*}
for a given target field $M \colon \mathbb{R}^2 \longrightarrow \mathrm{SO}(3)$. This easily reduces to the condition $M_{21}+\mathrm{i} M_{31} \neq  \pm \psi_1 = \pm e^{\mathrm{i}\hat{\psi}_1}$ for every point in space, which is directly analogous to the three-SLM case in beam generation. Thus, by a similar logic, the existence of a continuous decomposition reduces to the existence of a lift $\tau \colon \Sigma \longrightarrow S^1$, where $\Sigma = \{(x,y)\colon M_{11}(x,y)=0\}$ and $\tau = M_{21}\lvert_\Sigma + \mathrm{i} M_{31}\lvert_\Sigma$, and obstructions are completely characterized by the cohomology class
\begin{equation*}
    [\tau] \in \check{H}^1(\Sigma; \mathbb{Z}).
\end{equation*}